\documentclass[twocolumn]{aastex62}

\newcommand\aastex{AAS\TeX}

\def\Fermi{\textit{Fermi}}

\def\deg{\hbox{$^\circ$}}

\received{2017 November 8}
\accepted{2018 May 1}

\shorttitle{\aastex\ origins of the gamma-ray flux variations of NGC 1275}
\shortauthors{Tanada et al.}

\begin{document}

\title{The origins of the gamma-ray flux variations of NGC 1275 based on 8 years of $\Fermi$-LAT observations}
\correspondingauthor{Kazuhisa Tanada}
\email{arcmin\_phase19@fuji.waseda.jp}

\author{K. Tanada}
\affiliation{Research Institute for Science and Engineering, Waseda University, 3-4-1, Okubo, Shinjuku, Tokyo, 169-8555, Japan}

\author{J. Kataoka}
\affiliation{Research Institute for Science and Engineering, Waseda University, 3-4-1, Okubo, Shinjuku, Tokyo, 169-8555, Japan}

\author{M. Arimoto}
\affiliation{Research Institute for Science and Engineering, Waseda University, 3-4-1, Okubo, Shinjuku, Tokyo, 169-8555, Japan}
\affiliation{Faculty of Mathematics and Physics, Institute of Science and
Engineering, Kanazawa University, Kakuma, Kanazawa, Ishikawa 920-1192, Japan}

\author{M. Akita}
\affiliation{Research Institute for Science and Engineering, Waseda University, 3-4-1, Okubo, Shinjuku, Tokyo, 169-8555, Japan}

\author{C. C. Cheung}
\affiliation{Space Science Division, Naval Research Laboratory, Washington, DC 20375, USA}

\author{S. W. Digel}
\affiliation{W. W. Hansen Experimental Physics Laboratory, Kavli Institute for Particle Astrophysics and Cosmology, Department of Physics and
SLAC National Accelerator Laboratory, Stanford University, Stanford, CA 94305, USA}

\author{Y. Fukazawa}
\affiliation{Department of Physical Science, Hiroshima University, 1-3-1 Kagamiyama, Higashi-Hiroshima, Hiroshima 739-8526, Japan}

\begin{abstract}
We present an analysis of 8 years of $\Fermi$-LAT ($>0.1$ GeV) $\gamma$-ray data obtained for the radio galaxy NGC 1275. The $\gamma$-ray flux from NGC 1275 is highly variable on short ($\sim$ days to weeks) timescales, and has steadily increased over this 8-year timespan. By examining the changes in its flux and spectral shape in the LAT energy band over the entire dataset, we found that its spectral behavior changed around 2011 February {($\sim$ MJD 55600).} The $\gamma$-ray spectra at the early times evolve largely at high energies, while the photon indices were unchanged in the latter {times} despite rather large flux variations. To explain these observations, we suggest {that} the flux changes in the early times were caused by injection of high-energy electrons into the jet, while later, the $\gamma$-ray flares were caused by a changing Doppler factor owing to variations in the jet Lorentz factor and/or changes in the angle to our line of sight. To demonstrate the viability of these scenarios, we fit the broad-band spectral energy distribution data with a one-zone synchrotron self-Compton (SSC) model for flaring and quiescent intervals before and after 2011 February. To explain the $\gamma$-ray spectral behavior in the context of the SSC model, the maximum electron Lorentz factor would have changed in the early times, while a modest change in the Doppler factor adequately fits the quiescent and flaring state $\gamma$-ray spectra in the later times.

\end{abstract}






\section{Introduction} \label{sec:intro}
NGC 1275 is a well-known galaxy located at the center of the Perseus cluster, at a redshift of $z=0.0179$, with an active galactic nucleus (AGN) classified as a Seyfert 1.5.
It is one of a few non-blazar AGNs detected in both high-energy (HE; $>0.1$ GeV) and very high-energy (VHE; $>0.1$ TeV) $\gamma$ rays so far, and is the brightest radio galaxy at GeV energies  \citep[e.g.,][]{2010ApJ...720..912A, 2016arXiv161102986R}.
Analyzing long-term observations of such bright sources and studying differences between low flux (quiescent) and high flux (flaring) intervals could contribute to our understanding of the physical mechanisms responsible for the $\gamma$-ray emission of AGN.
{Generally however, the flux variations of various types of AGN, originating from different emission regions such as the accretion disk, disk coronae, and jets, are of the colored noise-type
\citep[e.g.,][and reference therein]{2017ApJ...837..127G}.}
For this reason, the distinction between quiescent and flaring states of AGN is always arbitrary;
however, we can approximately categorize them into two such states by using the difference of their flux in the $\gamma$-ray band.

NGC 1275 has been widely observed at different wavelengths from the radio band to VHE $\gamma$-ray band.
In the radio band, NGC 1275 hosts the exceptionally bright radio source Perseus A (also known as 3C 84) with a pair of radio jets and large-scale Fanaroff-Riley type I (FR-I) radio morphology. 3C 84 has been studied in detail with very long baseline interferometry (VLBI).
These observations reveal a compact core and jet components to the south that are moving steadily outwards at 0.3 milli-arcseconds per year \citep{2009AJ....138.1874L}.
The presence of a faint counter-jet implies a jet angle to our line of sight, $\theta$ = 30\deg--55\deg\ on milli-arcsecond scales \citep{1994ApJ...430L..41V,1994ApJ...430L..45W,2006PASJ...58..261A}, with 
lower estimates of $\theta \leq 14.4\deg$ for the smallest-scale structures \citep{1992A&A...260...33K}
Taking these constraints together indicates curvature of the jet away from the line of sight at larger scales \citep{2006MNRAS.366..758D}.
Curiously, a newer sub-pc scale component was discovered near the nucleus in 2007 with continuously increasing radio flux \citep{2012MNRAS.423L.122N}, {and even larger suggested jet angle to the line of sight of $\theta \sim 65\deg$ \citep{2017MNRAS.465L..94F}.}
The increase of the radio flux is considered to have originated in the activity of the {jets.}
Furthermore, recent studies reported that {increase} in the $\gamma$-ray {flux} may be correlated with the radio {flux densities} in NGC 1275 \citep{2014MNRAS.442.2048D,J. A. Hodgson}.
On short timescales {(days and weeks)}, the radio {flux densities were} not highly variable while the $\gamma$-ray {flux} varied widely, indicating that the $\gamma$ rays are produced closer to the core than the radio emission.
A systematic $\gamma$-ray study is therefore valuable to investigate the physical origin of the high-energy emission from either the jets, or closer to the central supermassive black hole.


In the optical band, photometric observations of the core exhibited some flares and hour-scale time variations \citep{1999A&A...351...21P}.
There are two scenarios ascribing the observed optical variability either to the accretion disk in the system \citep{1995A&A...296..628N}, or to the unresolved segment of the jets \citep{2000MNRAS.314..359H}.
However, the low optical polarization in the core, at the level of $\sim 0.4\%$ from Kanata observations \citep{2013PASJ...65...30Y}, indicates the jet (synchrotron) 
emission is not a major component in the optical band.
In contrast, \cite{2014A&A...564A...5A} reported that the optical core (KVA telescopes) and $\gamma$-ray (\Fermi \ Large Area Telescope, LAT) light curves from 2009 October to 2011 February are in good agreement at a $4-5~\sigma$ significance level, thereby suggesting that the $\gamma$-ray nonthermal continuum from NGC 1275 has the same origin as the optical emission.

The Perseus cluster is one of the brightest clusters in X-rays, with the 0.5--8 keV emission dominated by the thermal bremsstrahlung of the intracluster medium cooling flow \citep{2003ApJ...590..225C,2011MNRAS.418.2154F}.
Although $Swift$-BAT could not resolve the nucleus spatially, an excess of a nonthermal hard X-ray emission from the cluster central regions (galaxy NGC 1275) was detected in the 15--55 keV range \citep{2009ApJ...690..367A}.
A correlation between the variable X-ray (5--10 keV) and HE $\gamma$-ray fluxes was reported \citep{2016arXiv160803652F}, but the origin of the nuclear X-ray emission (i.e., disc/corona versus jets) is still under debate.

In high-energy $\gamma$ rays, NGC~1275 was discovered with the {$\Fermi$-LAT}, with an overall spectral energy distribution (SED) from the radio to VHE band {consistent with the standard} synchrotron self-Compton (SSC) jet model \citep{2009ApJ...699...31A,2010ApJ...715..554K}. Together with simultaneous MAGIC VHE observations (from 2009 to 2011), its $\gamma$-ray spectrum from 100 MeV to 650 GeV can be well fit either by a log-parabola or by a power-law function with a sub-exponential cutoff. {The applied SSC model indicates} a relatively small $\theta = 10$\deg--15\deg, and a jet bulk Lorentz factor, $\Gamma_{\rm b} \sim 10$ \citep{2014A&A...564A...5A}. These physical parameters indicate that NGC 1275 is a misaligned BL Lac object \citep{2017arXiv170407960X}.
However, the results were obtained from an analysis over a relatively small time interval ($\sim$ 1.4 years). Because NGC 1275 exhibits different $\gamma$-ray flux states, the estimates of the jet physical parameters in  the SSC model (i.e., Doppler factor, electron spectrum) in different activity states can help in gaining an understanding of the changing temporal and spectral behaviors of NGC 1275.

In this study, we investigate the $\gamma$-ray emission from NGC 1275 with the increased photon statistics of 8 years of $\Fermi$-LAT observations.
These data allow us to study the spectral properties, variations in flux and photon index, and the distribution of the highest-energy photons.
In particular, the $\gamma$-ray flux states are well characterized in the flux and spectral hardness (i.e., hardness ratio, photon index) plane, in which
NGC 1275 is known to exhibit different radiation states \citep{2014MNRAS.442.2048D}.
The $\Fermi$-LAT long-term observational data help us to understand the transitions between these radiation states.

We present the details of the $\Fermi$-LAT analysis and data reduction in Section \ref{sec:observations}.
Our analysis results are presented in Section \ref{sec:results} and discussed in Section \ref{sec:discussion}, 
based on fitting the overall SED data of NGC 1275 with the one-zone SSC model. Our conclusions are presented in Section \ref{sec:conclusion}.
{In this paper we assume a standard flat $\Lambda \rm CDM$ cosmology with $H_0=70 \ \rm km \ s^{-1} \ Mpc^{-1}$ and $\Omega_m=0.29$ \citep{2014ApJ...794..135B}. This corresponds to a linear scale of 1 arcsec$ \ =\  $360 pc at a luminosity distance of $D_L$ =76.7 Mpc.}
Throughout this paper, the errors correspond to $1~\sigma$ confidence level.

\newpage
\section{Observations and Analysis} \label{sec:observations}

\subsection{$\Fermi$-LAT Observations} \label{sec:fermi}

The LAT is a pair-conversion telescope onboard the $\Fermi$ spacecraft, which was launched in 2008, and is designed to cover the energy band from 20 MeV to greater than 300 GeV. 
The LAT has a large effective area ($\sim$ 9000 cm$^2$ on axis at 10 GeV) and a large field of view ($\sim$ 2.4 sr).
The 68\% containment radius for $E > 10$ GeV is approximated as $\theta_{68} = 0\fdg15$, and is approximated as $\theta_{68} = 3\fdg5$ for $E = 100$ MeV.  A detailed description of the detector is provided in \cite{2009ApJ...697.1071A}.

The 8 years of data used in this study comprise {spacecraft data obtained in sky-survey mode} between 2008 August 4 and 2016 November 15 (MJD 54683 and 57707, respectively).
We applied a zenith angle cut of 90\deg\ to reduce the contamination due to $\gamma$ rays from the Earth's limb. The same zenith cut is considered in the exposure calculation using the \Fermi\ Science Tool {\tt gtltcube}\footnote{The \Fermi\ Science Tools and standard diffuse emission models are available from the $\Fermi$ Science Support Center, \url{http://fermi.gsfc.nasa.gov/ssc}}. We used the recommended ``Source" class events \citep{2012ApJS..203....4A} appropriate for a standard analysis. The lower energy bound was fixed at 100 MeV, and the region of interest (ROI) radius was fixed at 30\deg\ in this study to consider the tails of the LAT point-spread function (PSF) sufficiently. The data were analyzed using the $\Fermi$ Science Tools version v10r0p5, and Instrument Response Functions (IRFs) {P8R2\_SOURCE\_V6\footnote{\url{https://fermi.gsfc.nasa.gov/ssc/data/analysis/software/}}}. The Pass 8 data provide considerable improvements over the data used in earlier \Fermi-LAT studies, with enhancements in direction reconstruction and classification of events, better energy measurements, and significantly increased effective area allowing us to study the $\gamma$-ray emission from NGC 1275 more precisely.


To investigate the $\gamma$-ray flux variations of NGC 1275 (RA=49\fdg951, DEC=41\fdg512; J2000),  
we calculated the light curve and photon index variation by using a binned {\tt gtlike} {analysis} (standard maximum-likelihood spectral estimator provided with the $\Fermi$ science tools) using 14-day time bins. 
For simplicity, we fit the data with a single power-law function. The definition is provided in Section~\ref{sec:8-yearSPEC}, Eq.~\ref{eq:PLbestfit}.
As the diffuse background emission should not be variable, we fixed the parameters of the Galactic (gll\_iem\_v06.fits) and isotropic diffuse (iso\_P8R2\_SOURCE\_V6\_v06.txt) templates to their maximum likelihood values for the entire 8-year data set. In addition, we let the normalization and the index of NGC 1275 free.
We considered the $\gamma$-ray point sources listed in the 3rd \Fermi-LAT source catalog (3FGL) \citep{2015ApJS..218...23A} within 30\deg\ of NGC 1275 {(3FGL J0319.8+4130)}.
{We also considered a new point source (RA=48\fdg321, DEC=41\fdg526; J2000) found by generating a residual Test Statistic \citep[TS;][]{1996ApJ...461..396M} map over a $10\deg \times 10\deg$ region centered on NGC 1275 using the {\tt gttsmap} tool.}
While the spectral parameters for the point sources within 10\deg\ of NGC 1275 were left free in the fits, we fixed the parameters for the point sources beyond 10\deg\ to the maximum likelihood values for the 8-year data set.
Figure~\ref{fig:LC_2weeks} shows the variation of the flux and photon index and the TS
value against time.
For every time bin, the obtained TS values exceed 40 (corresponding to $\sim 6~\sigma$).
We note that the effects of the systematic uncertainties are not included in the error bars, and we estimate them to be on the order of 5\% based on the systematic uncertainty of the effective area\footnote{See \url{ https://fermi.gsfc.nasa.gov/ssc/data/analysis/LAT_caveats.html}}. In addition, the systematic uncertainties are not independent between time bins in the analysis.


\subsection{$NuSTAR$ Observations} \label{sec:nustar}
The $NuSTAR$ Observatory consists of two co-aligned telescopes focusing hard X-rays in the 3--79 keV range onto two focal plane modules, FPMA and FPMB \citep{2013ApJ...770..103H}. It provides relatively low-background imaging capabilities (18\arcsec \ full-width half-maximum) in the hard X-ray band with 2 $\mu$sec relative timing resolution.
NGC 1275 was observed by $NuSTAR$ on 2015 November 3 starting at 03:21:08 UT {in target-of-opportunity mode.} The effective on-source exposure was 20.0 ksec.
The FPMA and FPMB data were re-processed following the standard $NuSTAR$ data analysis system pipeline, nupipeline v0.4.5 and HEASoft (v6.19), together with the $NuSTAR$ calibration files from CALDB version 20160502.
 
In this paper, we extracted the 3-79 keV spectrum of NGC 1275 using a region with a radius of 10\arcsec \ and evaluated {the local background in an annulus around the source with the inner and outer radii of 10\arcsec and 30\arcsec, respectively.}
The resultant {\it NuSTAR} spectrum of NGC 1275 is shown in Section~\ref{SSCmodel}.

\subsection{$Chandra$ Observations} \label{sec:chandra}
{The Advanced CCD Imaging Spectrometer (ACIS-I) detector on board the $Chandra$ X-ray Observatory has an angular resolution of $\sim 0.5\arcsec$ on-axis operating in the range of 0.2--10 keV.
Its very high resolution allows us to investigate the nonthermal emission from the vicinity of the core.
To avoid the effects of pileup, we selected three observations (ObsId 12025, 12033, 12036; PI Fabian) with large offset angles $> 7.5 \arcmin$ from the nucleus \citep{2014A&A...564A...5A}.
The exposure times are 18.2 (on 2009 November 25), 19.1 (on 2009 November 27) and 48.5 ksec (on 2009 December 2), respectively.
The data were analyzed using the $Chandra$ Interactive Analysis of Observations (CIAO) software v4.8, and the $Chandra$ CALDB v4.7.4. Spectral analysis was performed using XSPEC v12.9.

We extracted the 0.5-9.5 keV spectrum of NGC 1275 using a region with a radius of 5\arcsec \ and evaluated the local background in an annulus around the source with the inner and outer radii of 5\arcsec and 15\arcsec, respectively.
The merged three spectra were fitted simultaneously by adopting the model $phabs \times (mekal + zphabs \times powerlaw)$ in XSPEC, where $phabs$ and $zphabs$ correspond to the Galactic and internal photoelectric absorptions, $mekal$ represents the thermal emission from the hot diffuse gas, and $powerlaw$ is the nonthermal power-law emission from the unresolved central core.
We fixed the hydrogen column density from the Galaxy to $1.5 \times 10^{21} \ \rm cm^{-2}$ \citep{2013PASJ...65...30Y},
and the internal absorption column density of $(1.4 \pm 0.3) \times 10^{21} \ \rm cm^{-2}$ is obtained from the fit.
The metal abundance and hydrogen density of $mekal$ were fixed to 0.7 solar value and $0.1 \ \rm cm^{-3}$, respectively \citep{2014A&A...564A...5A}. We therefore obtained the temperature of $14.4 \pm 3.3 \ \rm keV$.
The photon index was $2.11 \pm 0.16$, and the total integral flux in the  2--10 keV band was $1.14 \times 10^{-11} \ \rm erg \ cm^{-2} \ s^{-1}$.
The resultant power-law component of the {\it Chandra} X-ray spectrum (bow-tie) of NGC 1275 is shown in Section~\ref{SSCmodel}.

}

\section{Results} \label{sec:results}
\subsection{Light curve}\label{sec:lightcurve}



\begin{figure}[ht!]
  \centering
  \includegraphics[scale=0.21, bb=58 0 1268 848]{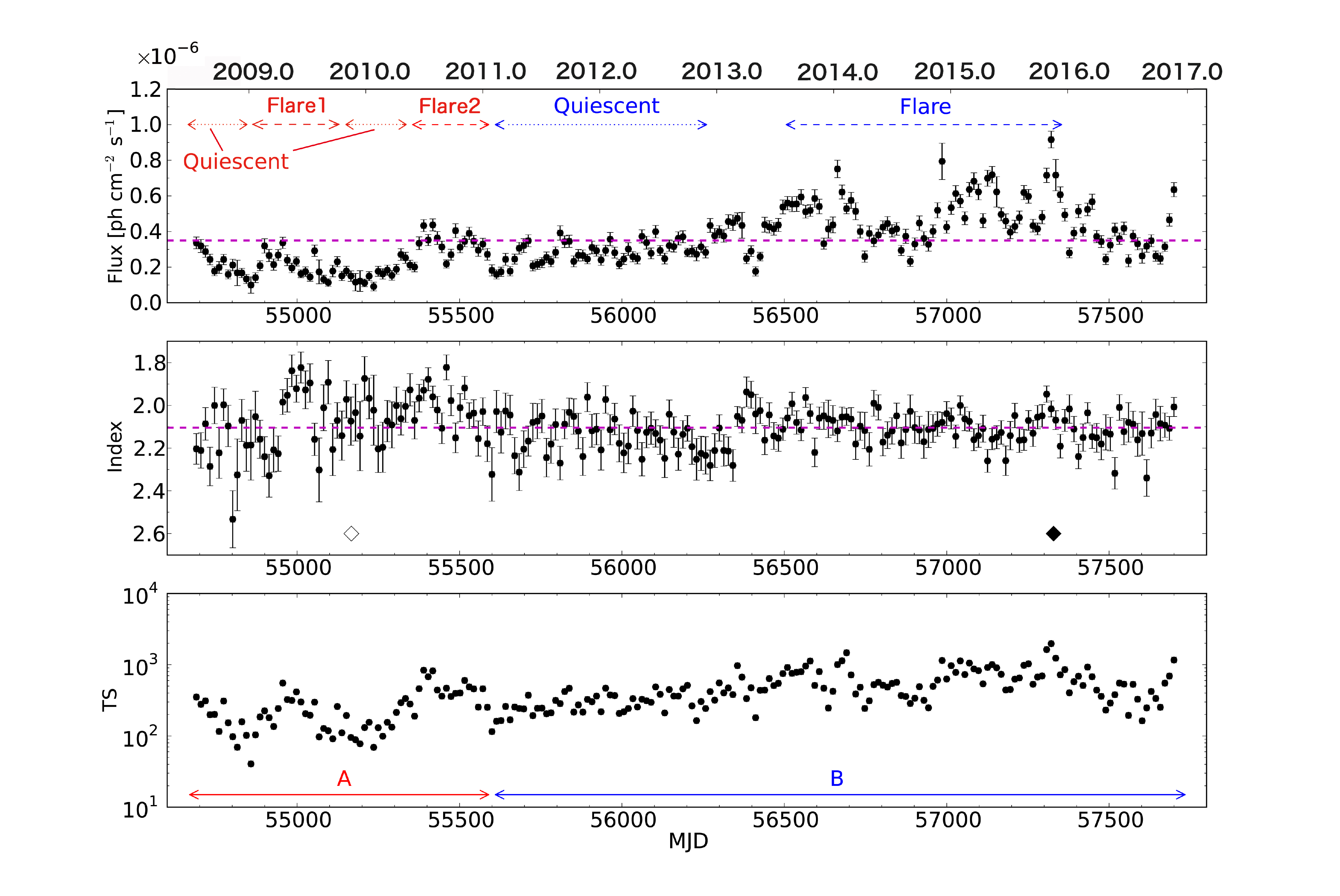}
  \caption{$\Fermi$-LAT light curve and variation of photon index for NGC 1275 over the time interval from 2008 August - 2016 November in 2-week time-bins. $\bf{Top \ panel:}$ changes in the $E > 100$ MeV flux. {The magenta dashed line shows the average flux derived from the 8 year analysis.} $\bf{Middle \ panel:}$ changes in the power-law photon index. {The magenta dashed line shows the average photon index.} $\bf{Bottom \ panel:}$ changes in the {Test Statistic}. According to the differences in the {photon index behaviors during the flaring intervals,}
we have divided the light curve into two time intervals showing a {large variation}
(epoch A) and 
{no significant variation} (epoch B). 
In addition, we defined quiescent and the flaring intervals within each epoch as indicated by dotted and dashed double-headed arrows, respectively.
{The open and filled diamonds in the middle panel represent the time of the $Chandra$ and $NuSTAR$ observations, respectively.}
}
  \label{fig:LC_2weeks}
\end{figure}

As can be seen in Figure~\ref{fig:LC_2weeks}, the $\gamma$-ray flux increases gradually by a factor of eight from MJD 55200 to MJD 57300.
{We divided the light curve into intervals with relatively low-flux (quiescent interval) and high-flux (flaring interval) states -- 
i.e., quiescent intervals (MJD  54683--54865, 55061--55369, 55607--56278) and flaring intervals (``flare 1": MJD 54865--55061, ``flare 2": MJD 55369--55607, ``flare 3": MJD  56503--57371).}
To verify the validity of this separation {between the flaring and quiescent intervals}, we used the two-sample Kolmogorov--Smirnov test for the flux data,
which measures the probability that a univariate dataset is drawn from the same parent population as the other dataset.
The calculated probabilities that the quiescent and flaring flux distributions are the same are $0.04\%$ in epoch A and $0.0008\%$ in epoch B.

{Then, we divided the light curve into epoch A and epoch B according to the {photon index behaviors during the flaring intervals.}
While the photon index changes largely during the flaring intervals at the earliest times (MJD $<$ 55607  = epoch ``A" in Figure~\ref{fig:LC_2weeks}), there is no apparent change in the photon index during later times (MJD $>$ 55607 = epoch ``B" in Figure~\ref{fig:LC_2weeks}) despite the presence of larger $\gamma$-ray flares.
{In fact, the variance value of the photon index of epoch A ($0.020 \pm 0.004$) is larger than that of epoch B ($0.007 \pm 0.001$) at $\sim 3 \ \sigma$ significance level.}
Regarding the quiescent intervals, we can assume that the states of epoch A and epoch B are almost the same. It is divided just for convenience to calculate the difference spectra between quiescent and flaring intervals (in Section~\ref{sec:resolvedSPEC} and Section~\ref{SSCmodel}).}

\subsection{Spectral analysis}\label{sec:spectral}
\subsubsection{Eight-year accumulated spectrum}\label{sec:8-yearSPEC}

We used a binned likelihood analysis with {\tt gtlike} to investigate NGC 1275's average $\gamma$-ray spectrum for the 8-year LAT data set. 
We first fitted the $\gamma$-ray emission with {a single power-law function}

\begin{equation}
	\frac{dN}{dE} = N_0 \left( \frac{E}{100~\rm MeV} \right) ^{-\Gamma}.
	\label{eq:PLbestfit}
\end{equation}

\noindent The Galactic and isotropic diffuse background components are assumed to exhibit a power-law spectrum as well, and we allow their normalizations to be free. The maximum likelihood power-law parameters obtained from a binned {\tt gtlike} analysis were $N_{\rm 0} = (3.82\pm 0.04) \times 10^{-9} ~\rm {ph~cm^{-2}\,s^{-1}\,MeV^{-1}}$ at 100 MeV and $\Gamma = 2.10 \pm 0.01$, with a corresponding average flux of $F_{>100 \rm MeV} = (3.48 \pm 0.02) \times 10 ^{-7}\rm {ph~cm ^{-2}s ^{-1}}$ (only statistical uncertainties are considered throughout).

We then obtained a $\gamma$-ray spectrum by running {\tt gtlike} separately in 22 equally-spaced logarithmic energy bands from 100 MeV to 204.8 GeV. Because the significance was low (TS = 9.3)  in the 204.8 to 300.0 GeV band, we calculated a 2 $\sigma$ upper limit. In the analysis, the normalizations of the diffuse backgrounds were fixed to their maximum likelihood values from the entire 8-year data set. The energy range of the 8-year LAT spectrum extends to slightly higher energies than in previous works, which indicated significant emission up to 102.4 GeV \citep{2010ApJ...715..554K,2011MNRAS.413.2785B}.


\begin{figure}[ht!]
  \centering
  \includegraphics[scale=0.45, bb=20 0 567 386]{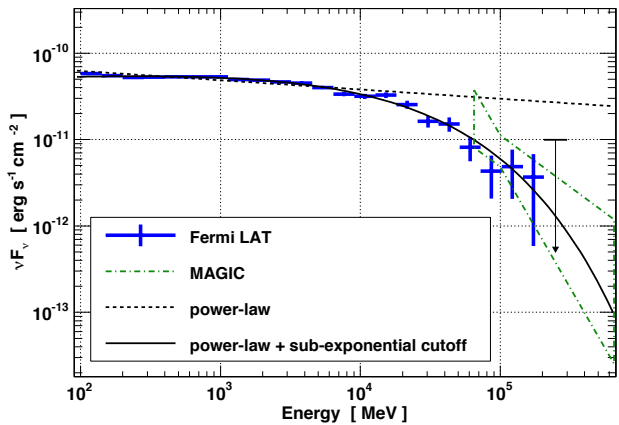}
  \caption{Average 8-year  $E > 100$ MeV LAT spectrum of NGC 1275 from 2008 August 4 to 2016 November 15. The dashed line indicates the power-law function determined from {\tt gtlike} while the solid line indicates the best-fit power law with a sub-exponential cutoff. The MAGIC spectrum (from 2009 October to 2010 February) is represented by green dashed bow-tie \citep{2014A&A...564A...5A}.}
  \label{fig:Cnu}
\end{figure}

Figure~\ref{fig:Cnu} clearly indicates a cutoff in the spectrum. Thus, we refit the data with {a sub-exponentially cutoff power-law function}
\begin{equation}
	\frac{dN}{dE} = N_0 \left(\frac{E}{100~\rm {MeV}} \right) ^{-\Gamma} \exp \left(- \sqrt{\frac{E}{E_{\rm c}}} \right).
	\label{eq:PLEXPbestfit}
\end{equation}

\noindent The maximum likelihood parameters were $N_{\rm 0} = (3.69 \pm 0.04) \times 10^{-9} ~\rm {ph~cm^{-2}\,s^{-1}\,MeV^{-1}}$ at 100 MeV, $\Gamma = 1.93 \pm 0.01$, $E_{\rm c} = 12.0 \pm 1.7$ GeV, and an average flux, $F_{>100 \rm MeV} = (3.34 \pm 0.03) \times 10 ^{-7}\rm {ph~cm^{-2}\,s^{-1}}$.
The Test Statistic is TS = 89986 corresponding to a formal significance of $\sim~300~\sigma$.
By comparing the log-likelihood \citep{1996ApJ...461..396M} with a single power-law function, we can conclude that 
{a sub-exponentially cutoff power-law function provides a better representation of the data than a single power-law with a significance of $\sqrt{\Delta \rm TS} \sim~13~\sigma$
 ($\Delta$TS = 2 $\times$ ($\log{L_{\rm plexpcut}} - \log{L_{\rm pl}}$) = 169.8,} where $L_{\rm plexpcut}$ and $L_{\rm pl}$ are the likelihood of the sub-exponentially cutoff power-law function and single power-law function, respectively.).
Additionally, the obtained best-fit parameters are within $\sim~1~\sigma$ of the MAGIC VHE data up to 650 GeV \citep{2014A&A...564A...5A}, as shown in Figure~\ref{fig:Cnu}.

\subsubsection{Time-resolved spectrum}\label{sec:resolvedSPEC}

Next, we calculated the time-resolved spectrum {divided into 10 equally-spaced logarithmic energy bands from 100 MeV to 102.4 GeV} for each of the intervals we defined in Section~\ref{sec:lightcurve}.
{According to the modeling results presented in Section~\ref{sec:8-yearSPEC}, we assume that their spectra can well be represented by sub-exponentially cutoff power-law functions.}
The best-fit parameters are shown in Table~\ref{table:eachspec}, 
along with the significances compared with {single-power-law fits.}

\begin{figure*}[]
 \begin{minipage}[b]{0.5\linewidth}
  \centering
  \includegraphics[scale=0.72, bb=20 20 360 252]{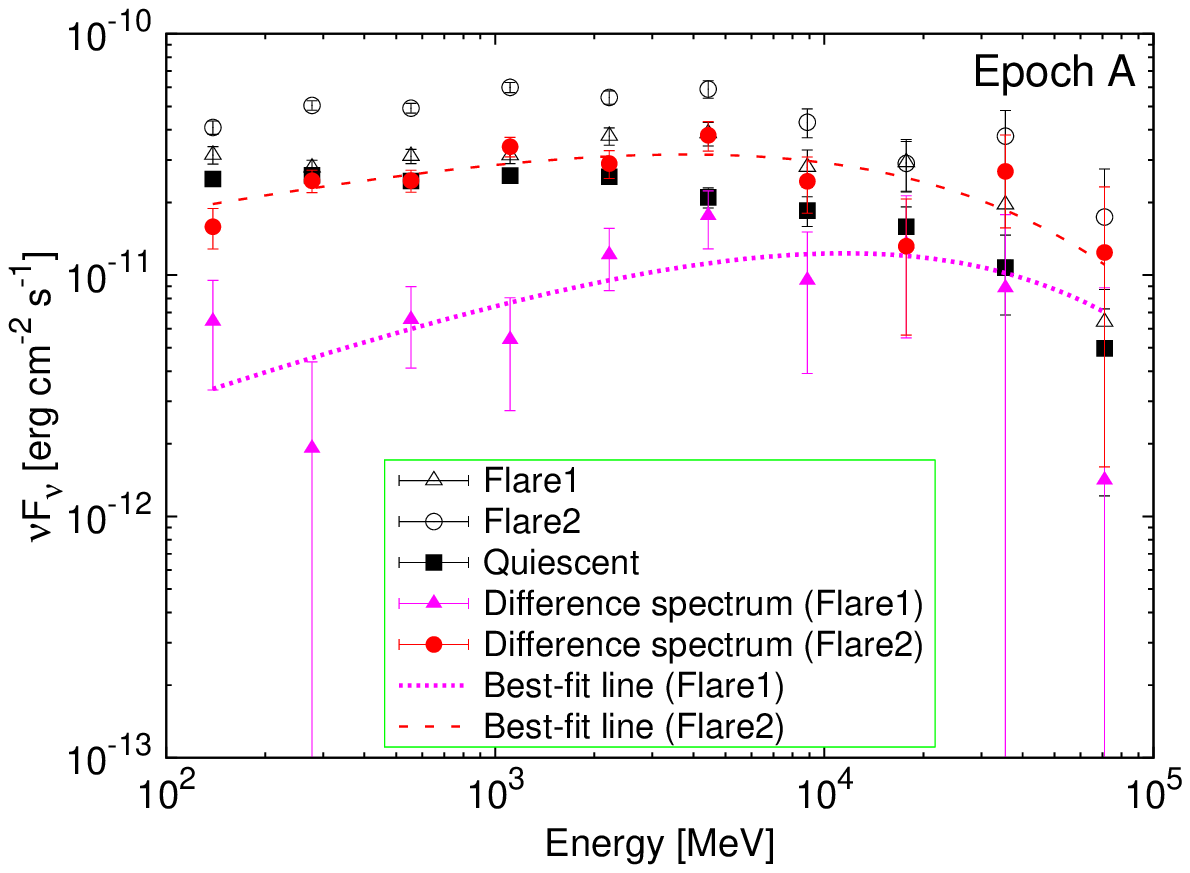}
  \label{subfig:A}
 \end{minipage}
 \begin{minipage}[b]{0.5\linewidth}
  \centering
  \includegraphics[scale=0.72, bb=20 20 360 252]{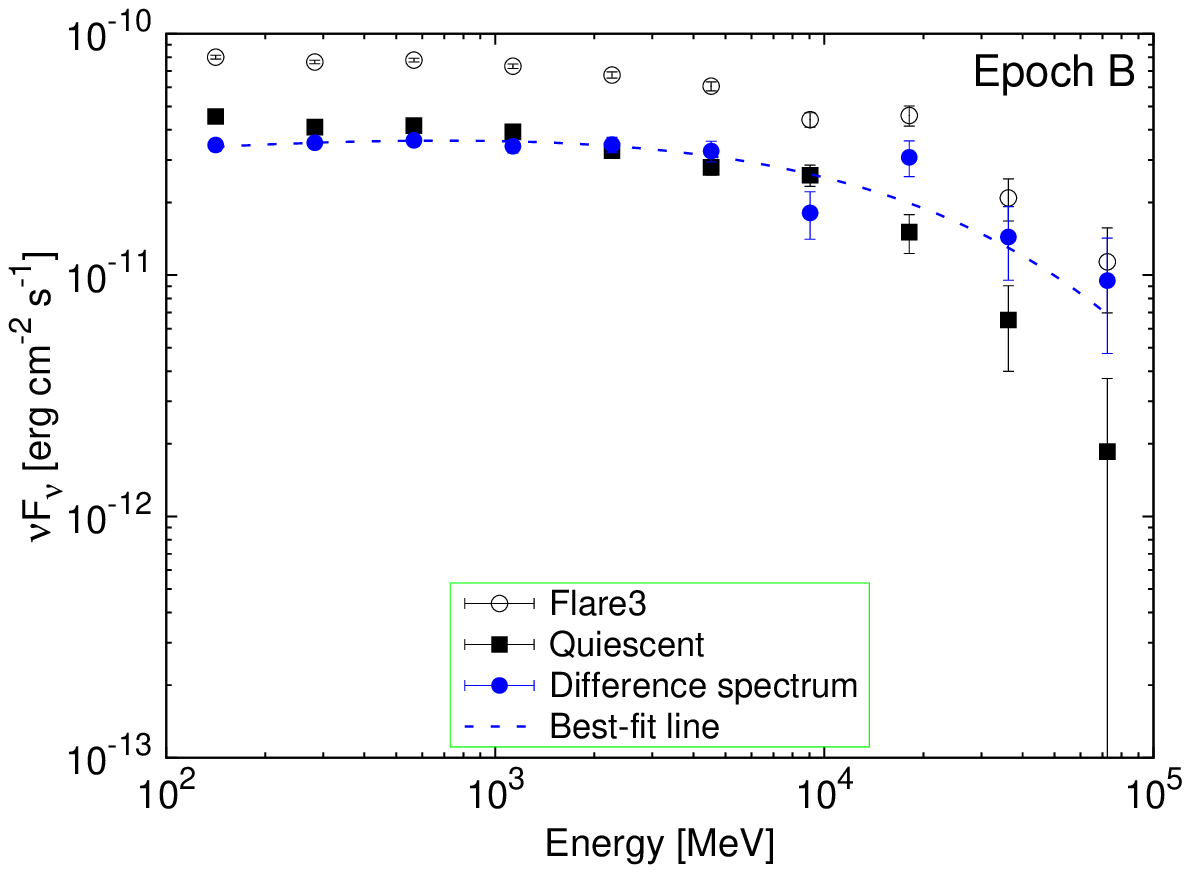}
  \label{subfig:B}
 \end{minipage}

 \caption{{Spectra and difference spectra} ($\nu F_{\nu}$), calculated as the difference between the flaring and quiescent {data points}. {The flaring states are plotted as open triangles (flare 1 in epoch A) and open circles (flare 2 in epoch A, flare 3 in epoch B).
The quiescent states in both epochs are plotted as filled squares. $\bf{Left \ panel:}$ difference spectra in epoch A (flare 1: magenta filled triangles, flare 2: red filled circles) and their best-fit lines (flare 1: magenta dotted line, flare 2: red dashed line) $\bf{Right \ panel:}$ difference spectrum in epoch B (blue  filled circles) and its best-fit line (blue dashed line).}}
 \label{fig:subtracted_SED}
\end{figure*}

Moreover, to extract representative flaring state spectra, we calculated differences between the spectra for each epoch as the difference between the flaring and quiescent data points. 
Figure~\ref{fig:subtracted_SED} shows the SED of the quiescent intervals and flaring intervals in epoch A (left panel) and epoch B (right panel).
{The obtained photon index of the best-fit power law with a sub-exponential cutoff function for each difference spectrum is shown in Table~\ref{table:eachspec}, where the cutoff energy were fixed to their maximum likelihood values from the entire 8-year data set.}
The difference spectra in epoch A  are harder than that in epoch B, which indicates that a hard spectral component is injected during the flares in epoch A.
On the other hand, the {photon index} of the difference spectrum in epoch B is almost the same as the {photon index} before subtraction. These results suggest that the physical origins of the flares {in epoch A and epoch B} are different.

\begin{deluxetable*}{ccccccc}[]
\tablecaption{Best-fit parameters for the  defined sub-intervals}
\tablecolumns{7}
\tablewidth{0pt}
\tablehead{
\colhead{Epoch} &
\colhead{State} &
\colhead{$N_0 ~[10^{-9} ~\rm {ph~cm ^{-2} s ^{-1} MeV ^{-1}}$]} &
\colhead{$\Gamma$} &
\colhead{$E_{\rm c}~\rm[GeV]$} &
\colhead{Significance\tablenotemark{a}~[$\rm \sigma$]} &
\colhead{{$\Gamma$ (Difference spectrum)\tablenotemark{b}}}
}
\startdata
A & Quiescent & $1.89 \pm 0.08$ & $1.94 \pm 0.04$ & $18 \pm 10$ & 4.2 & $-$ \\ 
   & Flare 1 & $2.14 \pm 0.13$ & $1.87 \pm 0.05$ & $20 \pm 13$ & 3.4 & $1.51 \pm 0.10$ \\ %
   & Flare 2 & $3.06 \pm 0.09$ & $1.79 \pm 0.04$ & $11 \pm 4$ & 6.1 & $1.72 \pm 0.04$ \\ \hline
B & Quiescent & $2.69 \pm 0.05$ & $1.93 \pm 0.03$ & $8 \pm 2$ & 7.8 & $-$ \\
   & Flare 3 & $5.48 \pm 0.09$ & $1.93 \pm 0.02$ & $12 \pm 3$ & 11.1 & $1.88 \pm 0.02$ \\
\enddata
\tablenotetext{a}{Significance of the sub-exponentially cutoff power-law function compared with a single power-law function.}
\tablenotetext{b}{{The photon index of the best-fit power law with a sub-exponential cut-off function for each difference spectrum.}}
\tablecomments{The parameters of the best-fit power law with a sub-exponential cut-off function for each time-resolved spectrum are obtained by {\tt gtlike}. The definitions of the parameters are in Eq.~\ref{eq:PLEXPbestfit}.}
\label{table:eachspec}
\end{deluxetable*}

\subsection{Angular separation of $\gamma$-ray photons}
\label{sec:EMAX}
\begin{figure}[ht!]
  \centering
  \includegraphics[scale=0.72, bb=10 0 360 252]{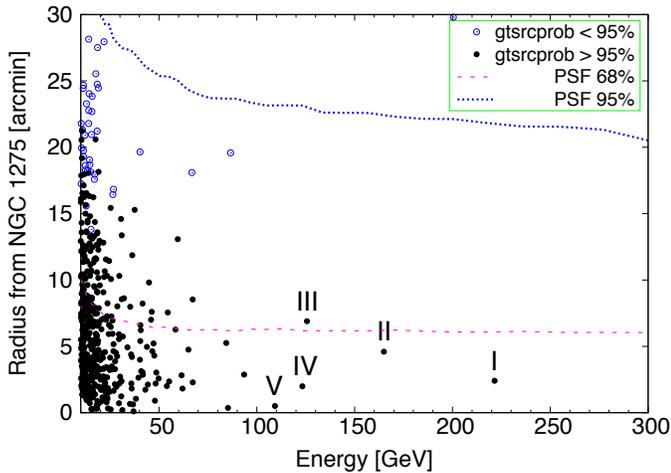}
  \caption{Angular separation of $\gamma$-ray photons against NGC 1275 as a function of photon energy ($E > 10$ GeV). The black filled circles and blue open circles describe the photons which have probabilities greater than 95\% or less than 95\% calculated by {\tt gtsrcprob}, respectively. In addition, the LAT 68\% and 95\% containment radii are indicated with a dotted magenta and blue lines, respectively. We note that the PSF curves represent averages over the field of view. The roman numerals label the five highest-energy photons in order.}
  \label{fig:EMAX}
\end{figure}

To examine whether the highest-energy photons detected by $\Fermi$-LAT near NGC 1275 are associated with the galaxy,
we {investigated} the angular separations of the individual $\gamma$ rays from NGC 1275, as shown in Figure~\ref{fig:EMAX}. 
We calculated probabilities that the detected photons are associated with NGC 1275 using the $\Fermi$ Science tool, {\tt gtsrcprob}, which indicated that the highest-energy photons {are likely} associated with NGC 1275. {To run {\tt gtsrcprob}, we performed an unbinned likelihood analysis here.} The contribution of {IC 310, which is a point-like VHE $\gamma$-ray emitter, is considered} to be small because {it lies $\sim 0.6^{\circ}$ from NGC 1275 \citep{2016A&A...589A..33A}.}

\begin{figure}[ht!]
  \centering
  \includegraphics[scale=0.75, bb=10 0 360 252]{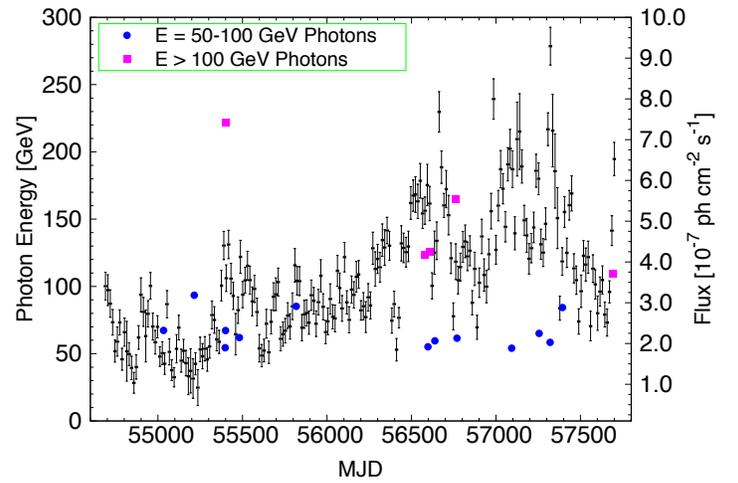}
  \caption{Arrival times and energies of high-energy photons plotted on the 8-year $E > 100$ MeV LAT light curve in 2-week bins ({black points}). The left vertical axis indicates the photon energy, while the right vertical axis indicates $\gamma$-ray flux. The blue circles indicate photons with energies {from 50 GeV to 100 GeV}. The magenta squares indicate photons with energies greater than 100 GeV (see Table \ref{table:EMAX}).}
  \label{fig:LC_50GeV}
\end{figure}

The five highest-energy photons with measured energies greater than 100 GeV, with 95\% probabilities of being associated with NGC 1275, are plotted in Figure~\ref{fig:EMAX}, with details provided in Table~\ref{table:EMAX}. 
Although \cite{2010ApJ...715..554K} reported that the highest-energy photon detected was 67.4 GeV during the first year of LAT observations ({MJD 54683--55061}), according to our analysis, the highest energy of the detected photons is 222 GeV during the 8-year interval considered here.
Moreover, we plot the high-energy photons with energies greater than 50 GeV on the 8-year light curve, as shown in Figure~\ref{fig:LC_50GeV}.
{This suggests that the arrival times of these high-energy photons are almost consistent with the flare intervals.}
{In particular,} the highest-energy photon of 222 GeV was detected during the flare 2 interval in epoch A, which might imply that the electrons in the jet were accelerated to higher energies during this interval.

\begin{deluxetable*}{cccccc}[htb]
\tablecaption{Details of the five highest-energy LAT photons}
\tablecolumns{6}
\tablewidth{0pt}
\tablehead{
\colhead{ } &
\colhead{Energy [GeV]} &
\colhead{Time [MJD]} &
\colhead{RA [\deg]\tablenotemark{a}} &
\colhead{DEC [\deg]\tablenotemark{a}} &
\colhead{Probability}
}
\startdata
  	I & 221.5 & $55402.4$ & 49.92 & 41.49 & 0.997\\ 
	$\rm I \hspace{-.1 em} \rm I$ & 164.0 & 56760.9 & 50.03 & 41.50 & 0.997\\
	$\rm I \hspace{-.1 em} \rm I \hspace{-.1 em} \rm I$ & 125.6 & 56610.8 & 50.06 & 41.48 & 0.994\\
	$\rm I \hspace{-.1 em} \rm V$ & 123.3 & 56578.0 & 49.98 & 41.53 & 0.999\\
	V & 109.2 & 57694.7 & 49.94 & 41.51 & 0.999\\
\enddata
\tablenotetext{a}{J2000.}
\tablecomments{The {Roman} numerals in the first column correspond to the order of the photon energies, as shown in Figure~\ref{fig:EMAX}. The probabilities being associated with NGC 1275 were calculated using the $\Fermi$ science tool {\tt gtsrcprob}.}
\label{table:EMAX}
\end{deluxetable*}

\section{Discussion} \label{sec:discussion}
\subsection{Hardness ratio}

\begin{figure}[ht!]
  \centering
  \includegraphics[scale=0.75, bb=10 0 360 252]{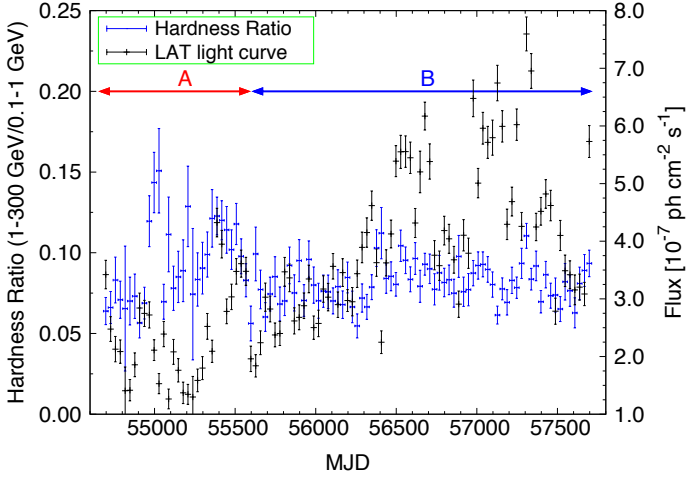}
  \caption{Hardness ratio defined as 1--300 GeV flux divided by 0.1--1 GeV flux against time for NGC 1275, in {1-month bins} (blue points). For comparison, we plotted the light curve with 1-month bin as well ({black points}).}
  \label{fig:HR_30days}
\end{figure}

To investigate the physical origins of the $\gamma$-ray flux increase in NGC 1275 for each epoch,
we calculated the hardness ratio (HR) defined as the LAT measured 1--300 GeV flux divided by 0.1--1 GeV flux.
Figure~\ref{fig:HR_30days} shows the HR in monthly bins and the flux light curve for comparison.
{The HR plot indicates that the spectral shape in epoch A changes considerably with relatively large variations in flux,
while the HR values do} not change significantly in epoch B even during the large flares.

{In Figure~\ref{fig:HR_vs_flux}, we show the {relation} between $\gamma$-ray HR and flux for the large flaring intervals in epoch A and B.}
{We obtained {the Spearman's rank correlation coefficients of -0.60 for the flare 1, 0.31 for the flare 2, and 0.01 for the flare 3; the corresponding chance probabilities of no correlation are 0.02, 0.24 and 0.92, respectively.}}
{In epoch A, the HR value changes significantly during the flaring intervals.}
{Interestingly, the spectrum became hard (HR $\sim 0.15$) a few months after the flux peak of the flare 1 interval. {This time lag might be explained by the gradual acceleration of the injected soft spectral component in the jet.}
On the other hand, in epoch B, the HR value does not change significantly during the flare 3 interval even though the flux change is very large.
Considering the difference between the slopes of the lines that fit the hardness ratio-flux points, we can suggest that these defined epochs represent different emission states.

\begin{figure}[ht!]
  \centering
  \includegraphics[scale=0.72, bb=10 0 360 252]{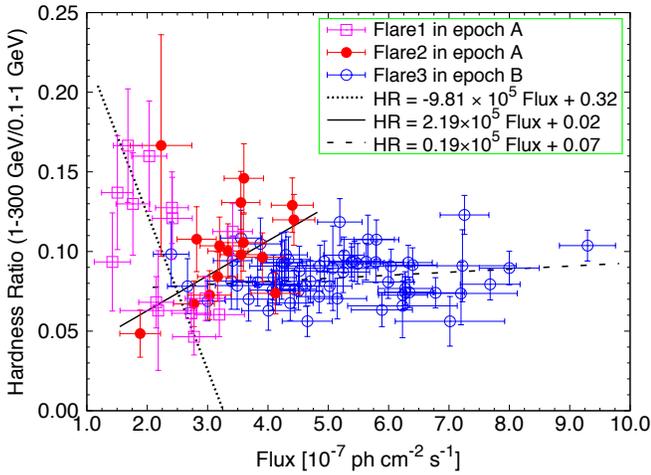}
  \caption{Hardness ratio against {flux} for NGC 1275, with {2-week bins}. 
{The data shown correspond to the three different flaring intervals, and lines denote the corresponding best-fit linear fits.}
}
  \label{fig:HR_vs_flux}
\end{figure}

We assume two scenarios to explain the temporal and spectral behaviors in epochs A and B.
{For epoch A, the variations of high-energy components can be interpreted as due to injections of freshly accelerated high-energy electrons into the emission zone.} 
In this case, the electrons are accelerated by some mechanism such as an internal shock in the jet \citep{1978MNRAS.184P..61R}.
Meanwhile, the flux variations in epoch B are due to changes in the Doppler factor ($\delta$) of a moving blob and/or changes in the electron density of the radiation zone.
When the Doppler factor changes, we can observe the spectral peaks with different energies ($\nu \propto \delta$), time variability ($t_{\rm var} \propto \delta ^{-1}$), and luminosities ($L_{\rm obs} \propto \delta ^ 4$) because of relativistic beaming. According to this scenario, the photon index does not change when the flux increases.

\subsection{Fractional variability}
{In order to characterize in more detail the difference in the source spectral variability between epochs A and B}, we evaluated fractional variabilities of the $\gamma$-ray light curves in different energy bands: 100--178 MeV, 178--316 MeV, 316--562 MeV, 562--1000 MeV, 1--3.16 GeV, 3.16--10 GeV, 10--54.8 GeV, and 54.8--300 GeV.
{First, in each energy band we calculated the excess variance $\sigma_{\rm rms}^2$, which is the net variance obtained after subtracting the noise variance from the total variance \citep[e.g.,][and reference therein]{2002ApJ...572..762Z}:}

\begin{equation}
	\sigma_{\rm rms}^2 = \frac{1}{N \overline{x}} \sum_{i=1}^{N} [(x_i -\overline{x}) ^2 - \sigma_i ^2] 
	= \frac{1}{\overline{x}^2} (\sigma_{\rm tot}^2 - \sigma_{\rm noise}^2),
	\label{eq:Fvar}
\end{equation}
where $x_i$ is the flux for the $i$-th bin in the light curve and $\overline{x}$ is the mean of $x_i$.
{The error estimate on $\sigma_{\rm rms}^2$ is $s_{\rm D}/(\overline{x}^2 \sqrt{N})$, where $s_{\rm D}$ is given by}

\begin{equation}
	s_{\rm D}^2 = \frac{1}{N-1} \sum_{i=1}^{N} \{ [(x_i -\overline{x}) ^2 - \sigma_i ^2] - \sigma_{\rm rms}^2 \overline{x}^2 \}^2.
	\label{eq:sD}
\end{equation}
Finally, we obtained the fractional variability {parameters in each energy band, $F_{\rm var} = \sqrt{\sigma_{\rm rms}^2}$, as shown in Figure~\ref{fig:Fvar} for the two epochs A and B.}

\begin{figure}[ht!]
  \centering
  \includegraphics[scale=0.72, bb=15 0 360 252]{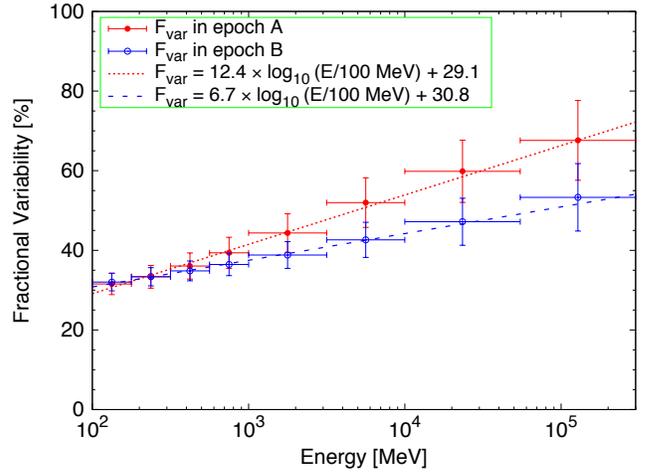}
  \caption{Energy dependence of variability of NGC 1275 with the red filled circles for epoch A and the blue open circles for epoch B. The variability parameter, i.e., the fractional variability ($F_{\rm var}$), was calculated for the eight different energy bands. {The red dotted and blue dashed lines are the best-fit logarithmic functions for each epoch.}}
  \label{fig:Fvar}
\end{figure}

{The best-fit line for each epoch in Figure~\ref{fig:Fvar} is calculated by fitting the data with a logarithmic function.
The obtained best-fit slopes of epochs A and B are $12.4 \pm 2.2$ and $6.7 \pm 1.7$, respectively, and they are different at $\sim~3~\sigma$ significance level.}
At higher energies, the values of $F_{\rm var}$ in epoch A are greater than that in epoch B. 
{This implies that the $\gamma$-ray continuum of the source in epoch A varies more strongly in the high-energy band, consistently with the idea that the origin of the flux increase in this epoch is related to changes (hardening) in the underlying electron energy distribution due to an enhanced acceleration of particles.}
{In epoch B, we do observe an elevated fractional variability in the high-energy band as well, albeit here the difference between the low- and high-energy segments of the $\gamma$-ray continuum is much less significant in this respect, in {agreement} with the idea that in this epoch at least the bulk of the observed flux {variations} is due to the changes in the jet Doppler factor, rather than an enhanced particle acceleration.}


\subsection{Synchrotron self-Compton model fits}
\label{SSCmodel}

{The overall double-peaked SED of the radio galaxy NGC 1275 is similar to that of a blazar, and in particular a low-power blazar of the BL Lac type} \citep{2009ApJ...699...31A,2014A&A...564A...5A}.
{Therefore, we attempt to model it with a standard homogeneous one-zone SSC model developed and widely used for BL Lac objects in general} \citep[for details see][]{1996ApJ...463..555I,2003ApJ...593..667M}.
The SSC model considers an electron energy density distribution with a form of $N(\gamma) = K \gamma^{-n} {(1+\gamma/\gamma_{\rm brk})^{-1}}$ for an electron Lorentz factor $\gamma$ ($\gamma_{\rm min} < \gamma < \gamma_{\rm max}$), 
where $\gamma_{\rm brk}$ represents the electron break Lorentz factor at which the radiative cooling time equals the acceleration time.
The electron density and the electron spectrum slope are represented by $K$ and $n$, respectively.
In addition, the other physical parameters of this model are the source radius, $R$, the magnetic field, $B$, and the Doppler factor, $\delta = 1/[\Gamma_{\rm b} (1-\beta \cos \theta)]$, where $\beta$ is the bulk speed of the plasma moving along the jet, the bulk Lorentz factor, $\Gamma_{\rm b} = [1-\beta^2]^{-1/2}$, and $\theta$ is the angle between the {jet axis and the line of sight}.
The source size $R$ can be approximately constrained by {the observed variability (flux doubling) timescale in the LAT band, $t_{\rm var} \simeq$ a few months, to be $R < c t_{\rm var} \delta \lesssim 10^{18}$ cm for the expected $\delta \leq 10$.}
The magnetic field $B$ can be obtained from the ratio of the synchrotron and SSC luminosities, $L_{\rm sync}/L_{\rm SSC} \simeq U_B/U_{\rm rad}$, where $U_B = B^2 /8 \pi$ is the magnetic energy density and $U_{\rm rad}$ is the synchrotron radiation energy density  \citep{1979rpa..book.....R}.

\subsubsection{Epoch A}
The left panel of Figure~\ref{fig:SSC} shows the multi-wavelength $\nu F_{\nu}$ SED of NGC 1275 obtained with the radio-to-high-energy $\gamma$-ray data including the LAT data for the {quiescent intervals (MJD 54683--54865, 55061--55369) and the flaring intervals (flare 1: MJD  54865--55061, flare 2: MJD 55369--55607)} {described in Section~\ref{sec:resolvedSPEC}}.
In the radio band, RATAN 600 \citep{2009ApJ...699...31A}, MOJAVE \citep{2009ApJ...699...31A}, and the archival NED (NASA/IPAC extragalactic database) data were used.
We used the same radio data in {all} of the quiescent and flaring intervals because the radio emission is considered to be considerably less variable than the $\gamma$-ray band \citep{2014MNRAS.442.2048D}.
In the optical/UV band, MITSuME \citep{2009ApJ...699...31A}, {\it Swift}-UVOT, and NED data were used.
As the optical emission is contaminated by the host galaxy, the optical data do not contribute to the SSC fitting. 
The RATAN 600, MOJAVE, and MITSuME data are contemporaneous with the LAT quiescent data in 2008, and the {\it Swift}-UVOT data {were} obtained from an observation in 2007.
The data in the X-ray band such as {\it Chandra} {(this work; MJD 55160--55167)} and {\it Swift}-BAT \citep{2009ApJ...690..367A} correspond to the quiescent state of NGC 1275.
The VHE data are derived from the MAGIC observations from 2009 to 2010 \citep{2014A&A...564A...5A}.

We fit the SED with the one-zone SSC model using the observational data from the quiescent and flaring $\gamma$-ray flux states.
{The overall trend of the SEDs is adequately represented by the one-zone SSC model both in the quiescent and flaring intervals as shown in Figure~\ref{fig:SSC}; however, detailed comparison suggests significant deviation between the data and model especially in the soft X-ray band. A similar discrepancy can be seen in Figure 7 of \cite{2009ApJ...699...31A}, and that paper therefore considered a more complicated, decelerating flow model \citep{2003ApJ...594L..27G} to obtain a better fit to the data. In reality, not only the velocity gradient but other physical parameters, such as magnetic field strength, jet cross section and even the electron spectrum itself may vary at the same time along the jet path \citep[see, for example,][]{1980ApJ...235..386M}. Nevertheless, the one-zone SSC model is a rough but useful way to consider the origin of the SED evolution without any complexities in the model \citep[e.g.,][]{1997A&A...320...19M}. Table \ref{table:SSCparam} reports the parameters obtained from our SSC model.}
The derived physical parameters for the quiescent interval such as the  
{source radius of $R = 0.8 \times 10^{18}$ cm, the magnetic field of $B = 0.04$ G, the electron density of $K \sim 45 \ \rm cm^{-3}$, the electron spectrum slope of $n = 2.6$ and the Doppler factor of $\delta = 2.7$ are almost the same as those for the flaring intervals}, 
and the values of the magnetic field are in the typical range found for BL Lacs \citep{2010MNRAS.401.1570T}.
However, we changed the {maximum Lorentz factor from $\gamma_{\rm max} = 2.5 \times 10^5$ (quiescent interval) to $\gamma_{\rm max} =  4.0 \times 10^5$ (flare 1) and $\gamma_{\rm max} =  3.5 \times 10^5$ (flare 2).} 
{We additionally changed the break Lorentz factor from $\gamma_{\rm brk} = 0.8 \times 10^5$ (quiescent interval) to $\gamma_{\rm brk} =  1.0 \times 10^5$ (flare 1) and $\gamma_{\rm brk} =  1.8 \times 10^5$ (flare 2).}
This SSC fitting indicates that the flux variation between the quiescent and flaring states is explained by changing only the electron Lorentz factor parameters that are related to the acceleration of electrons, which is consistent with our hypothesis for the $\gamma$-ray flux changes during epoch A.

The highest-energy photon of $\epsilon_{\rm max}=222$ GeV was detected {in the flare 2 interval} in epoch A as described in Section~\ref{sec:EMAX}. This photon is {considered to originate} from scattering in the Klein-Nishina (KN) regime, because the energy of the seed photon in the rest frame of the relativistic electron is larger than 511 keV.
As the energy of the scattered photon in the KN regime is provided by $\epsilon_{\rm max} \sim m_{\rm e} c^2 \gamma_{\rm max} \delta$ \citep{1998ApJ...509..608T}, we can estimate the maximum electron Lorentz factor to be $\gamma_{\rm max} \sim 1.6 \times 10^5$,
which is {smaller than}
the result from the SSC fitting {($\gamma_{\rm max} = 3.5 \times 10^5$ in the flare 2 interval).}
{However, we note that the MAGIC observation detected higher energy photons of $\sim 650$ GeV \citep{2014A&A...564A...5A}, which implies a larger maximum electron Lorentz factor.}

\subsubsection{Epoch B}

The LAT data for the quiescent interval (MJD  55607--56278) and the flaring interval {(flare 3: MJD  56503--57371)} are plotted in the right panel of Figure~\ref{fig:SSC}.
For epoch B, we also plot the {\it NuSTAR} data {on 2015 November 3} (this work), which corresponds to {MJD 57329 in} the $\gamma$-ray flaring interval in epoch B.
The other data plots are the same as described for epoch A.

We fit the SED with the one-zone SSC model to the emission of the $\gamma$-ray quiescent state and flaring state, 
and the best-fit parameter values are shown in Table \ref{table:SSCparam}.
Moreover, we can confirm that the SED data of NGC 1275 for epoch B are well represented by the one-zone SSC model, as shown in Figure~\ref{fig:SSC}.
In particular, the fit in the flaring interval is consistent with the {\it NuSTAR} data, 
which suggests that the X-ray variability component is the same as in the $\gamma$ ray (i.e., originating in the jet).
The derived physical parameters for epoch B such as the
magnetic field of $B = 0.04$ G, the electron spectrum slope of $n=2.6$ and 
the maximum electron Lorentz factor of $\gamma_{\rm max} = 1.0 \times 10^5$ (which is smaller than used to fit the data in epoch A) are unchanged between the quiescent and flaring intervals.
Meanwhile, the Doppler factor of the flaring interval is $\delta =3.6$, which is larger than that during the quiescent interval, $\delta = 2.7$. In addition, the electron density changed from $K = 48 \ \rm{cm^{-3}}$ (quiescent interval) to $K = 270 \ \rm{cm^{-3}}$ {(flare 3)}, and the source radius changed from $R = 1.0 \times 10^{18} ~\rm cm$ (quiescent interval) to $R = 0.4 \times 10^{18} ~\rm cm$ {(flare 3)}, which indicates that the physical parameters of the blobs such as the Doppler factor $\delta$ and the viewing angle $\theta$ in the jet for the two intervals are not the same.
{Interestingly, the overall SED data cannot be fitted solely by changing the bulk Lorentz factor from $\Gamma_{\rm b} = 2.0$ (quiescent interval) to $\Gamma_{\rm b} = 3.3$ {(flare 3)}.
It also requires the jet-viewing angle to be changed from $\theta= 20\deg$ (quiescent interval) to $\theta= 16\deg$ {(flare 3)}.}
This could indicate the direction of motion of the blob in the jet is closer to the line of sight when the flux increased.
{We can therefore assume}
that the bright $\gamma$ rays are emitted {in the proximity} of the core (milli-arcsecond scales){,} considering the VLBI observations \citep{1992A&A...260...33K,1994ApJ...430L..41V,1994ApJ...430L..45W,2006PASJ...58..261A}
suggest the jet angle to the line of sight decreases with proximity to the core. Thus, we likely observed different $\gamma$-ray {flux} because of changes in Doppler beaming {due to} changing locations of the emission region.

From the obtained SSC results, we can suggest that the origins of the flux variations of NGC 1275 are clearly different depending on the observation periods.
Although the jet-viewing angle parameters are small ($\theta \sim 20\deg$) in both epochs compared with $\theta = 30-55\deg$ obtained by VLBI radio observations, the fitting results {generally support the scenarios we presented here.}


\begin{figure*}[]
 \begin{minipage}[b]{0.5\linewidth}
  \centering
  \includegraphics[scale=0.72, bb=10 10 360 252]{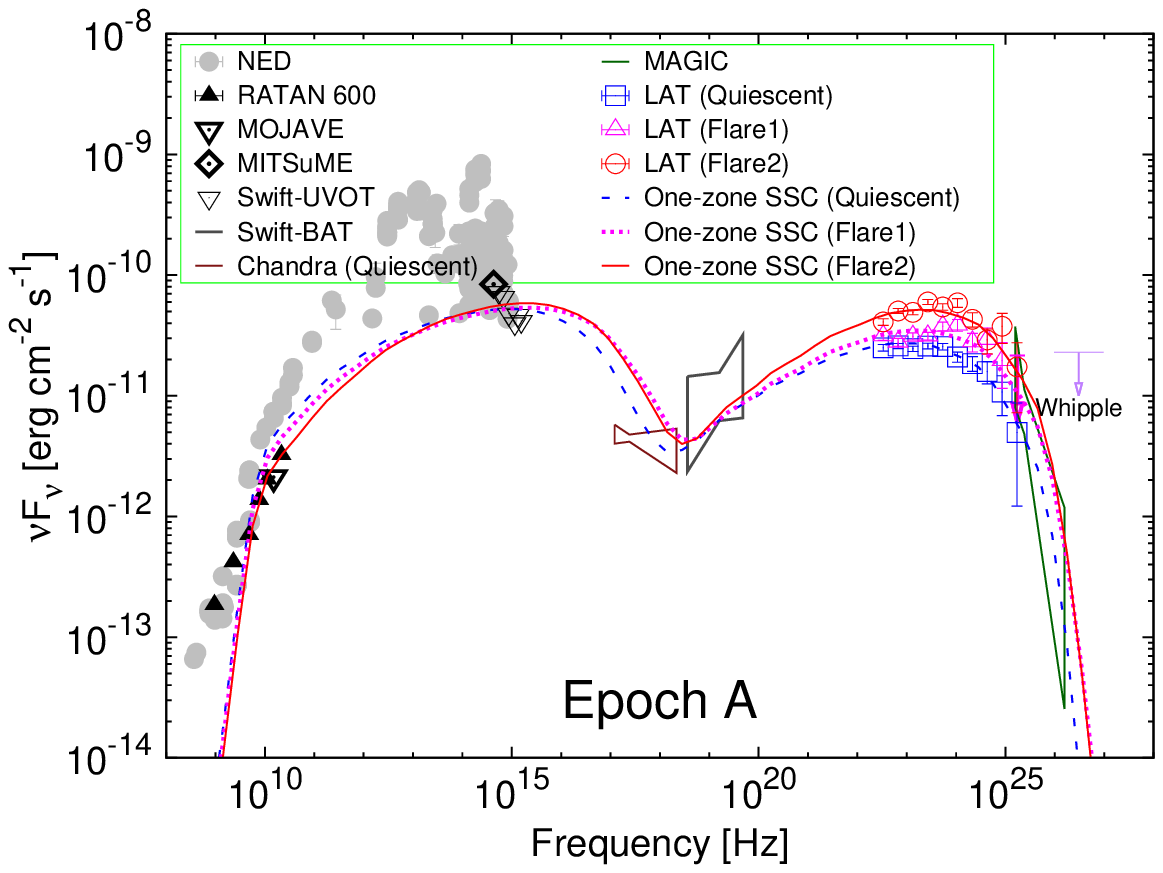}
  \label{subfig:A_SSC}
 \end{minipage}
 \begin{minipage}[b]{0.5\linewidth}
  \centering
  \includegraphics[scale=0.72, bb=10 10 360 252]{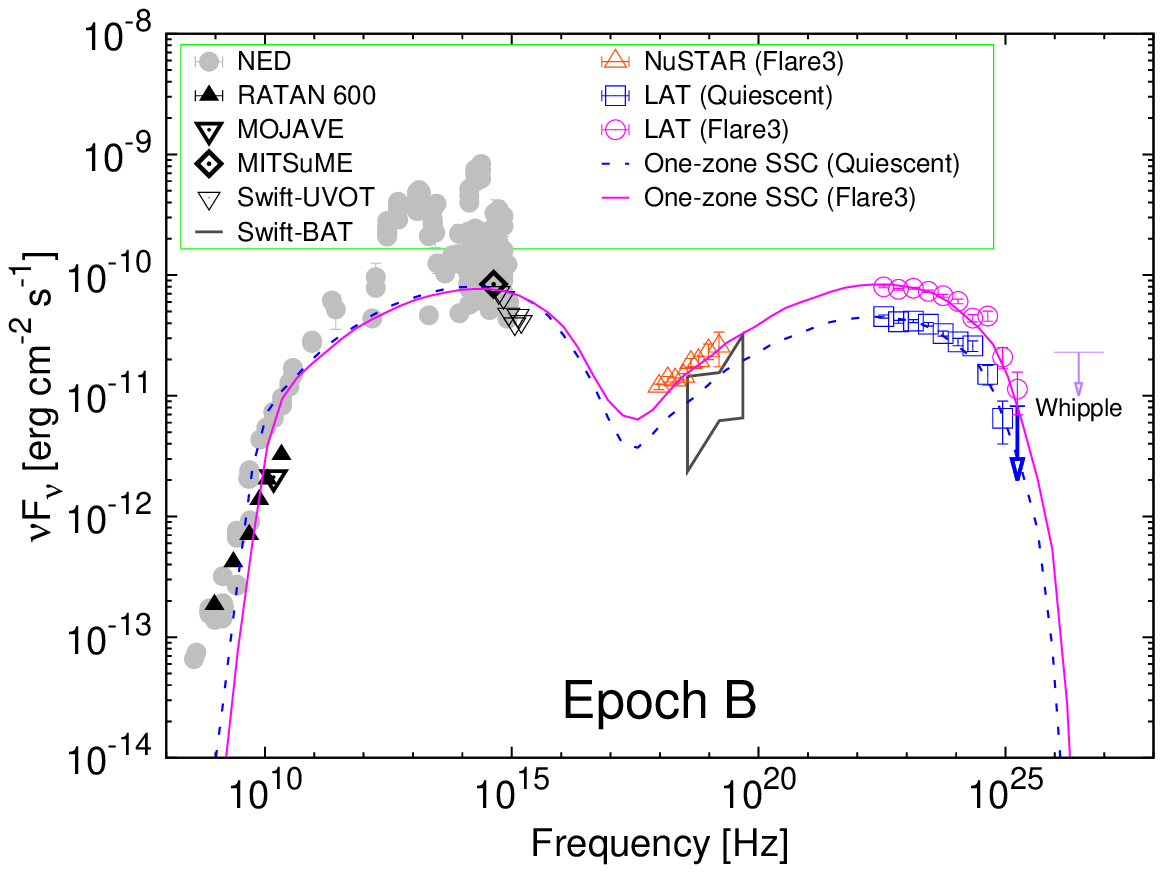}
  \label{subfig:B_SSC}
 \end{minipage}

 \caption{Overall SED of NGC 1275 obtained with multi-wavelength data, using RATAN-600 \citep{2009ApJ...699...31A}, MOJAVE \citep{2009ApJ...699...31A}, MITSuME \citep{2009ApJ...699...31A}, NASA/IPAC Extragalactic Database, {\it Swift}-UVOT \citep{2009ApJ...699...31A}, {\it NuSTAR} (this work), \Fermi-LAT (this work), {\it Chandra} bow-tie {(this work)}, {\it Swift}-BAT bow-tie \citep{2009ApJ...690..367A}, MAGIC bow-tie \citep{2014A&A...564A...5A}, and Whipple upper limit \citep{2006ApJ...644..148P}. 
{The quiescent and flaring SEDs are fitted with the one-zone SSC model and denoted by blue dashed line (quiescent intervals), magenta dotted line (flare 1), red solid line (flare 2), and magenta solid line (flare 3).}
 $\bf{Left \ panel:}$ the SSC model fitting for epoch A. {The blue squares, magenta open triangles, and red open circles represent the LAT data in the quiescent intervals (MJD 54683--54865, 55061--55369), flare 1 (MJD 54865--55061), and flare 2 (MJD 55369--55607), respectively.
{The {\it Chandra} data (brown bow-tie), which correspond to MJD 55160--55167 in the $\gamma$-ray quiescent interval, are also plotted.}
}
$\bf{Right \ panel:}$ the SSC model fitting for epoch B. The blue squares and magenta open circles represent the LAT data in the quiescent interval (MJD  55607--56278) and {flare 3} (MJD  56503--57371), respectively. The {\it NuSTAR} data (orange open triangles) which correspond to {MJD 57329 in the $\gamma$-ray flaring interval} are also plotted.}
 \label{fig:SSC}
\end{figure*}


\begin{deluxetable*}{cccccccccccc}[htb]
\tablecaption{{The fitted physical} parameters for the SSC model reported in Figure~\ref{fig:SSC}}
\tablecolumns{12}
\tablewidth{0pt}
\tablehead{
\colhead{Epoch} &
\colhead{State} &
\colhead{$R$ [cm]} &
\colhead{$B$ [G]} &
\colhead{$K$ $[\rm cm^{-3}]$} &
\colhead{$n$} &
\colhead{$\gamma_{\rm min}$} &
\colhead{$\gamma_{\rm brk}$} &
\colhead{$\gamma_{\rm max}$} &
\colhead{$\delta$} &
\colhead{$\Gamma_{\rm b}$} &
\colhead{$\theta$ $[^{\circ}]$}
}
\startdata
	A & Quiescent & $0.8 \times 10^{18}$ & 0.04 & 45 & 2.6 &10.0 & $0.8 \times 10^3$ & $2.5 \times 10^5$ & 2.7 & 2.0 & 20 \\
	& Flare 1 & $0.7 \times 10^{18}$ & 0.04 & 50 & 2.6 & 10.0 & $1.0 \times 10^3$ & $4.0 \times 10^5$ & 2.7 & 2.0 & 20\\
	 & Flare 2 & $0.6 \times 10^{18}$ & 0.04 & 50 & 2.6 & 10.0 & $1.8 \times 10^3$ & $3.5 \times 10^5$ & 2.7 & 2.0 & 20\\ \hline
	B & Quiescent & $1.0 \times 10^{18}$ & 0.04 & 48 & 2.6 &10.0 & $0.8 \times 10^3$ & $1.0 \times 10^5$ & 2.7 & 2.0 & 20\\
	 & Flare 3 & $0.4 \times 10^{18}$ & 0.04 & 270 & 2.6 &10.0 & $0.7 \times 10^3$ & $1.0 \times 10^5$ & 3.6 & 3.3 & 16\\
\enddata
\tablecomments{The obtained physical parameters for the quiescent and flaring states for both epochs A and B. The parameters are the source radius $R$, the magnetic field $B$, the electron density $K$, the electron spectrum slope $n$, the minimum electron Lorentz factor $\gamma_{\rm min}$, the break Lorentz factor $\gamma_{\rm brk}$, the maximum electron Lorentz factor $\gamma_{\rm max}$, the Doppler factor $\delta$, and the bulk Lorentz factor $\Gamma_{\rm b}$, the angle between the 
{jet axis and the line of sight} $\theta$.}
\label{table:SSCparam}
\end{deluxetable*}


\section{Conclusions} \label{sec:conclusion}

We presented an analysis of 8 years of $\Fermi$-LAT data for the nearby radio galaxy NGC 1275.
The LAT spectrum accumulated over 8 years is best described by a power law with {a sub-exponential cutoff}, with {the photon index $\Gamma = 1.93 \pm 0.01$ and the} cut-off energy $E_{\rm c} = 12.0 \pm 1.7$ GeV.
This is consistent with the result in the VHE band ($\sim$ 65--650 GeV) from MAGIC observations \citep{2014A&A...564A...5A}.
Based on positional coincidence, we found the highest-energy photon during the {8 years, with $E \sim 222$ GeV, has} $>99\%$ probability of association with NGC 1275. 

We analyzed the variations in the LAT lightcurve over the 8-year timespan and found that the correlation between the $\gamma$-ray flux and photon index changed around MJD  55607. 
In epoch A (MJD $<$ 55607), the emission from NGC 1275 is {interpreted as the injection of high-energy electrons in the jet}.
On the other hand, there is no apparent correlation in epoch B (MJD $>$ 55607) despite larger flares observed than in epoch A.
To explain these evidently different behaviors, we suggested different scenarios for the two epochs with the flux variations due to acceleration of the electrons during epoch A, and due to variations of the Doppler factor and/or the electron density during epoch B.
In order to verify these hypotheses, we fit the overall SED data with one-zone SSC models for flaring and quiescent time intervals during each epoch.
{The simultaneous observations of {\it Chandra} and {\it NuSTAR} can help us to obtain more accurate parameters.}
The SSC fitting for epoch A requires changing 
the maximum Lorentz factor from $\gamma_{\rm max} = 2.5 \times 10^5$ (quiescent interval) to {$\gamma_{\rm max} =  4.0 \times 10^5$ (flare 1) and $\gamma_{\rm max} =  3.5 \times 10^5$ (flare 2).}
Meanwhile, the flares in epoch B may be caused by variation of the Doppler factor from $\delta = 2.7$ (quiescent interval) to $\delta =3.6$ {(flare 3)}, which is interpreted as being due to changes of the bulk Lorentz factor and the angle between the blob velocity and the line of sight.
Although the jet-viewing angle parameter is small ($\theta \sim 20\deg$) in both epochs compared with the VLBI radio observations \citep{1994ApJ...430L..41V,1994ApJ...430L..45W,2006PASJ...58..261A,2017MNRAS.465L..94F}, the fitting results support our scenarios.
Particularly, for epoch B, the fitting requires a change of the jet-viewing angle from 20\deg\ (quiescent interval) to 16\deg\ {(flare 3)}, which indicates that the direction of motion of the blob in the jet is closer to the line of sight when the flux increases because of relativistic beaming effect.
The previous report that there is some curvature of the jet away from the core \citep{2006MNRAS.366..758D,2012ApJ...746..140S} supports this relationship between the jet-viewing angle and the flux increase.

Although we considered a scenario with only one emission zone in this study, the emission region and the radiation mechanism may be more complicated.
In fact, a few radio-emission jet components, known as C1 and C3 (moving to the south), exist near the nucleus \citep{2012MNRAS.423L.122N}. 
Moreover, \cite{J. A. Hodgson} found that some of the $\gamma$-ray emission likely originates in the C3 region, while short-time scale variability may be better correlated with the C1 mm-radio emission, suggesting multiple simultaneous sites of $\gamma$-ray emission within the same source.
Hence, we suggest that ultimately a multi-zone study might be justified, when more multi-wavelength data are considered. 
{
The multi-zone internal shock scenario involves sequential ejections of many blobs which have various emission region sizes, inducing multiple collisions at various distances from the core and a series of flares \citep{2001ApJ...560..659K,2001MNRAS.325.1559S,2003ApJ...584..153T}.
In particular, the different correlation between the HR and flux during the flares in epoch A may indicate multiple emission zones in the jet. The observed fluxes are then sums from the multiple zones; however its spectral behavior may change due to the emission from the different dominating regions where the electrons are injected.
}

In conclusion, we suggest that the origins of the flux variations of NGC 1275 are different for different epochs.
This result is derived from the analyzing the 8 years of the \Fermi-LAT data, which included both flaring and quiescent states of $\gamma$-ray emission.
It is possible that these findings are be applicable to other FR-I AGNs, and we will report on investigation of long-term observations for other objects in the future.

\acknowledgments

The \textit{Fermi} LAT Collaboration acknowledges generous ongoing support
from a number of agencies and institutes that have supported both the
development and the operation of the LAT as well as scientific data analysis.
These include the National Aeronautics and Space Administration and the
Department of Energy in the United States, the Commissariat \`a l'Energie Atomique
and the Centre National de la Recherche Scientifique / Institut National de Physique
Nucl\'eaire et de Physique des Particules in France, the Agenzia Spaziale Italiana
and the Istituto Nazionale di Fisica Nucleare in Italy, the Ministry of Education,
Culture, Sports, Science and Technology (MEXT), High Energy Accelerator Research
Organization (KEK) and Japan Aerospace Exploration Agency (JAXA) in Japan, and
the K.~A.~Wallenberg Foundation, the Swedish Research Council and the
Swedish National Space Board in Sweden.
 
Additional support for science analysis during the operations phase is gratefully
acknowledged from the Istituto Nazionale di Astrofisica in Italy and the Centre
National d'\'Etudes Spatiales in France. This work performed in part under DOE
Contract DE-AC02-76SF00515.

Work by C.C.C. at NRL is supported in part by NASA DPR S-15633-Y.

This work was supported by JSPS KAKENHI Grant Numbers JP17H06362 (M.A.).
M.A. acknowledges the support from JSPS Leading Initiative for
Excellent Young Researchers program.

\software
{
HEASoft (v6.19), 
Fermi Science Tools, 
nupipeline (v0.4.5)
}


\end{document}